\documentclass[preprint]{aastex}
\usepackage{epsfig}

\def\ee{{\it e}}
\def\pul{J0737-3039 }
\def\pulA{J0737-3039A }
\def\pulB{J0737-3039B }

\shorttitle{Evidence for a Low Mass Progenitor mass of pulsar
J0737-3039B} \shortauthors{Piran \& Shaviv}

\begin{document}

\title{On the Low Progenitor Mass of the Binary Pulsar J0737-3039B}
\author{Tsvi Piran and Nir J. Shaviv}
\affil{Racah Institute of Physics, Hebrew University of Jerusalem, Jerusalem 91904, Israel}

\begin{abstract}
The observed orbital parameters of the recently discovered binary
pulsar \pul are used to constrain the progenitor system. In
particular, the small observed eccentricity and the small inferred
peculiar velocity imply that during the formation of the second NS
in the system, a very small mass $\Delta M  \lesssim 0.15 M_\sun$,
was ejected. A progenitor more massive than 1.4$M_\odot$ is
unlikely and a progenitor more massive than 2.3$M_\odot$ is
practically ruled out.  We therefore argue that the companion
removed most of the progenitor envelope prior to its collapse. The
second collapse may have been a SN of Type Ib, or more probably,
of Type Ic. Alternatively, it could have been a bare collapse, an option 
which is kinematically favored. 
Future constraints
on the peculiar motion could solidify this constraint on the mass
loss. A future determination of the pulsars' spin vectors and
proper motion could allow us the complete determination of the
conditions during the second collapse.
\end{abstract}

\section{Introduction}

The discovery of the new binary pulsar \pul \citep{Burgay2003,
Lyne2004}  opened a new ideal general relativistic laboratory. The
eclipses of \pulA  beyond \pulB provide a superb way to explore
pulsar magnetospheres. The relatively short life time of this
system to gravitational radiation emission has lead to a revision
of the binary merger rate in the galaxy
\citep{Burgay2003,Kalogera2004}. We point out in this letter that
the orbital parameters of this system, and in particular its low
eccentricity and its  low peculiar velocity pose strong limits on
the origin of this binary system. We argue that the progenitor of
the second neutron star must have been very light and that the
formation of this neutron star must have involved only a minimal
mass loss.

The mass loss during the second explosion could easily disrupt the
system. A mass loss prior to the second core collapse (usually due to
a common envelope phase) or a kick velocity are usually invoked as
ways to prevent the disruption of the binary during the second
Supernova. Indeed there is evidence for kicks given to neutron
stars at birth \citep[see][]{vdH,Burrows}. However, a kick velocity could either keep the ellipticity small or
reduce the CM motion. That is to say, a kick which keeps the
ellipticity small will increase the CM motion and vice versa. Therefore,
the prior mass loss must have been very significant in this system.

The analysis described here is similar in spirit to that of
\cite{Wex00} who calculated the pre-SN characteristics of the
binary pulsar system B1913+16 given the kinematics of the system. 
Nevertheless, several difference between
the systems imply that both the analysis and the type of
conclusions will be different. For example, we cannot place a
limit on the direction of the natal kick relative to the pulsar
spin direction. We can however place a very strong limit on the
progenitor mass.

On the other hand, the analysis here is different from two previous analysis of \pul by \cite{DvdH04} and \cite{WK04}. In these analyses, the kick velocity of the neutron star was constrained given the center of mass kinematics of the system {\em and} evolutionary constraints. In particular, it was assumed that the pre-SN star was more massive than $2.1 M_\sun$ \citep{WK04} or $2.3 M_\sun$ \citep{DvdH04}, since evolutionary scenarios do not give NS's forming from less massive pre-SN Helium stars  in tight binaries. Here, we do not assume any evolutionary constraint, or for that matter, any assumption on the type of the progenitor. That is to say, we use here only kinematic constraints. Moreover,  we also consider the kick velocity of the system as a whole, based on a probability analysis for finding the system as close to the galactic plane as it is. 

A minor difference between the analysis here and that of \cite{DvdH04} and \cite{WK04} arises from the fact it is now known that the companion of \pulA is a pulsar as well, with a shorter characteristic spin down age. As a consequence, the last collapse in the system was more recent than assumed before. Nevertheless, the slow evolution of the orbital separation and eccentricity implies that this will only give rise to a small quantitative correction.

We discuss the properties of the binary pulsar in \S
\ref{sec:properties}. In \S \ref{sec:kick} we estimate the probability that the
binary system obtained different natal kicks. In \S \ref{sec:evol} we consider the
evolution of the system and the dynamics of formation of the
second neutron star. We discuss the conclusions and possible
caveats in \S \ref{sec:concl}.

\section{Pulsars' Properties}
\label{sec:properties}

The binary system \pul is composed of two pulsars denoted A and B.
The system was discovered during a pulsar search carried out using
a multibeam receiver at the Parkes 64-m radio telescope in New
South Whales. At first only \pulA was detected \citep{Burgay2003}.
Further observations revealed that the companion in this
system is also a pulsar \pulB \citep{Lyne2004}. We begin by
summarizing the relevant parameters of this system.
\begin{itemize}
    \item {\it Orbital Parameters:} The orbital period is 2.454 hours.
    The separation of the pulsars today is $8.8 \times 10^{10}$cm and the
    eccentricity is 0.087779(5).   This implies (see Fig. \ref{fig:evol})
    that the system will merge within 83~Myr due to emission of
    gravitational radiation. The system is seen almost edge on and
    its inclination is $87^\circ$.
    \item {\it Pulsar life times:} The
    pulsars' periods, $P_{A,B}$ are: 22.7 msec and 2.773 sec respectively.
    $P_{A,B}$ and the time derivatives $\dot
    P_{A,B}$ provide estimates for the life times of the pulsars
    with $t_A \sim 210$~Myr and $t_B \sim 50$~Myr. These
    observations also provide estimates for the magnetic fields of
    the two binaries: $B_A \sim 6.3 \times 10^9$G and  $B_B \sim 1.6 \times 10^{12}$G.
    \item {\it Position:} The system is $\sim 600$~pc away from Earth. Its
    Galactic coordinates are: L=245.2357 and B= -4.5048, which
    implies that it is within $50$~pc from the Galactic plane.
    \item {\it Masses:} The pulsars's masses are almost equal with:
    $m_A = 1.377(5)M_\odot$ and $M_B = 1.250(5) M_\odot$.
\end{itemize}

We assume in the following that the present dual pulsar system was
born about 50~Myrs ago when pulsar B was formed (as inferred from its
lifetime). We infer that prior to that, pulsar A was spun up by
accretion from its companion and that this spin up process stopped
200~Myr ago.

\section{Peculiar Motion}
\label{sec:kick}

One of the important observations which can be used to constrain
the system its observed location---about 50~pc from the galactic
plane. This is obtained given the dispersion measure distance of
$\sim0.6$~kpc and its galactic $b=-4.5^\circ$ latitude
\citep{Lyne2004}. This implies that it is unlikely that the system
attained a large peculiar motion during the core collapse. If it had,
it would have left the galactic plane unless, by chance, it was
kicked within the galactic plane. A large velocity would have made
the system unbound from the Galactic potential. It would have then
travelled, within its lifetime a distance of  $5 {\rm kpc}
(v/100{\rm km/sec}) (t/50{\rm Myr})$.

If the system was vertically bound then the motion would have been
periodic. The vertical angular speed is given by
\begin{equation}
\Omega_z^2 = 4 \pi G \rho_0.
\end{equation}
A typical value for the density in the disk, $\rho_0$, is $0.25
M_\odot$/pc$^3$ \citep{Bahcall92}.  This gives an orbital period
of $P_z \sim 50$~Myr. The typical velocity for an object observed
at $z_{obs}$ is $V_z \sim  2 \pi z_{obs} / P_z$. A distance of
50~pc would imply then that the vertical velocity is of the order
of 6 km/sec.

To quantify the probability, we perform a Monte Carlo simulation.
We assume that the progenitor distribution has a Gaussian
distribution in the amplitude $A_{z,0}$ of the vertical
oscillation, with a width $\sigma_z = 25, 50$ or $100$~pc. At
the moment of  core collapse, we assumed it has a phase $\varphi_0$
within its vertical motion, such that:
\begin{equation}
 z_0 = A_{z,0} \sin \varphi_0~~,~~v_{z,0} = {2 \pi A_{z,0} \over P_z} \cos \varphi_0.
\end{equation}

We have no particular priors on the kick velocity, we therefore
assume the center of mass kick velocity has a log-normal
distribution between 1 and 1000 km/s, and that the direction of
the kick is random. Given a random center of mass kick velocity
${\bf v}_{cm}$, we then calculate the new vertical velocity, and
vertical amplitude:
\begin{equation}
 v_{z,1} = v_{z,0} + v_{cm,z}~~,~~A_{z,1} = \sqrt{z_0^2 +
\left(v_{z,1} P_z \over 2 \pi\right)^2}
\end{equation}
The final vertical location of the binary will be given by:
\begin{equation}
z_1 = A_{z,1} \sin \varphi_1,
\end{equation}
where we assume that $\varphi_1$ is random. If we could have been
certain about the pulsar's age as well as the periodicity $P_z$,
we could have calculated it from the kick and $\varphi_0$, but
this is not the case.

\begin{figure}
\hskip -0.5cm \center{\epsfig{file=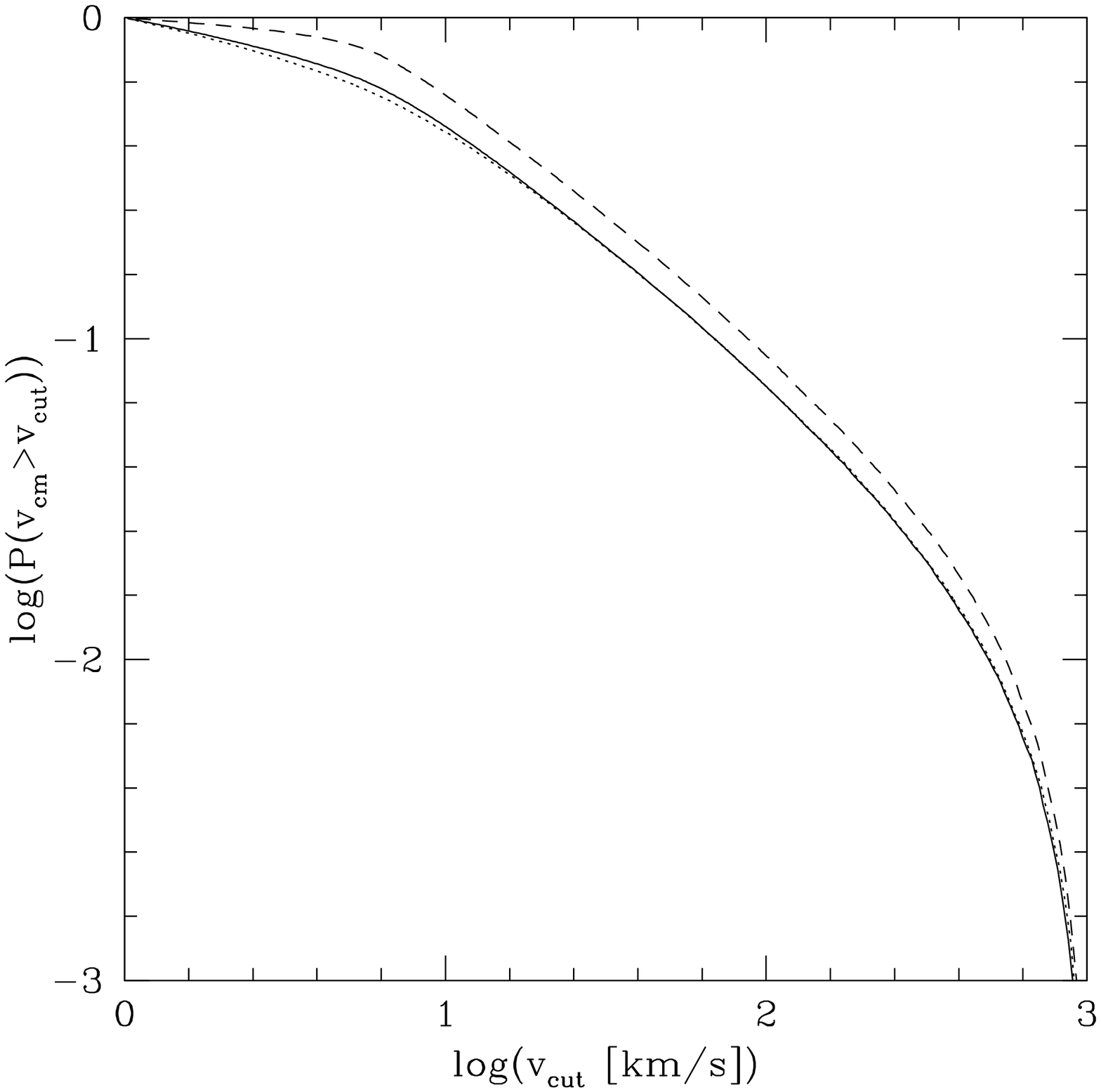,width=6in}} \caption{
\small \sf The probability that the binary attained a peculiar
motion given by $v_{cm}$ after the core collapse, that is larger
than a cutoff speed $v_{cut}$, given that today, it is located
roughly $\sim 50 pc$ from the galactic plane. The dashed, solid
and dotted lines assume that $\sigma_z=25,50,100$~km/s. The two
latter cases are virtually indistinguishable.
 }
 \label{fig:prob}
 \end{figure}

We repeat this procedure $10^7$ times, and collect all the events
which leave a binary located at $z_1 = z_{obs} = 50 \pm 5$~pc. We
then plot in Fig. \ref{fig:prob}  the distribution function
$P(v_{cm}>v_{cut})$. That is, we  plot the probability that the
given system had a CM kick larger than $v_{cut}$. We find that,
as expected, $v_{cm} \lesssim 15$~km/s at 68\% confidence, $v_{cm}
\lesssim 150$~km/sec, at 95\% confidence, and $v_{cm} \lesssim
500$~km/s at 99\% confidence. This velocity is small compared with
the internal circular velocity of the system $\sim 600$~km/s. This
will prove to be important.

The limit on the  vertical velocity and on one of the perpendicular
(to the line of sight) components of the velocity within the
Galactic disk can be found within a few years, as estimates of the
proper motion of the system will be obtained. Note that at a
distance of 600~pc, a velocity of 100~km/sec implies a proper motion
of 0.036\arcsec/yr.

\section{Evolution}
\label{sec:evol}  
\begin{figure}
\hskip -0.5cm
\center{\epsfig{file=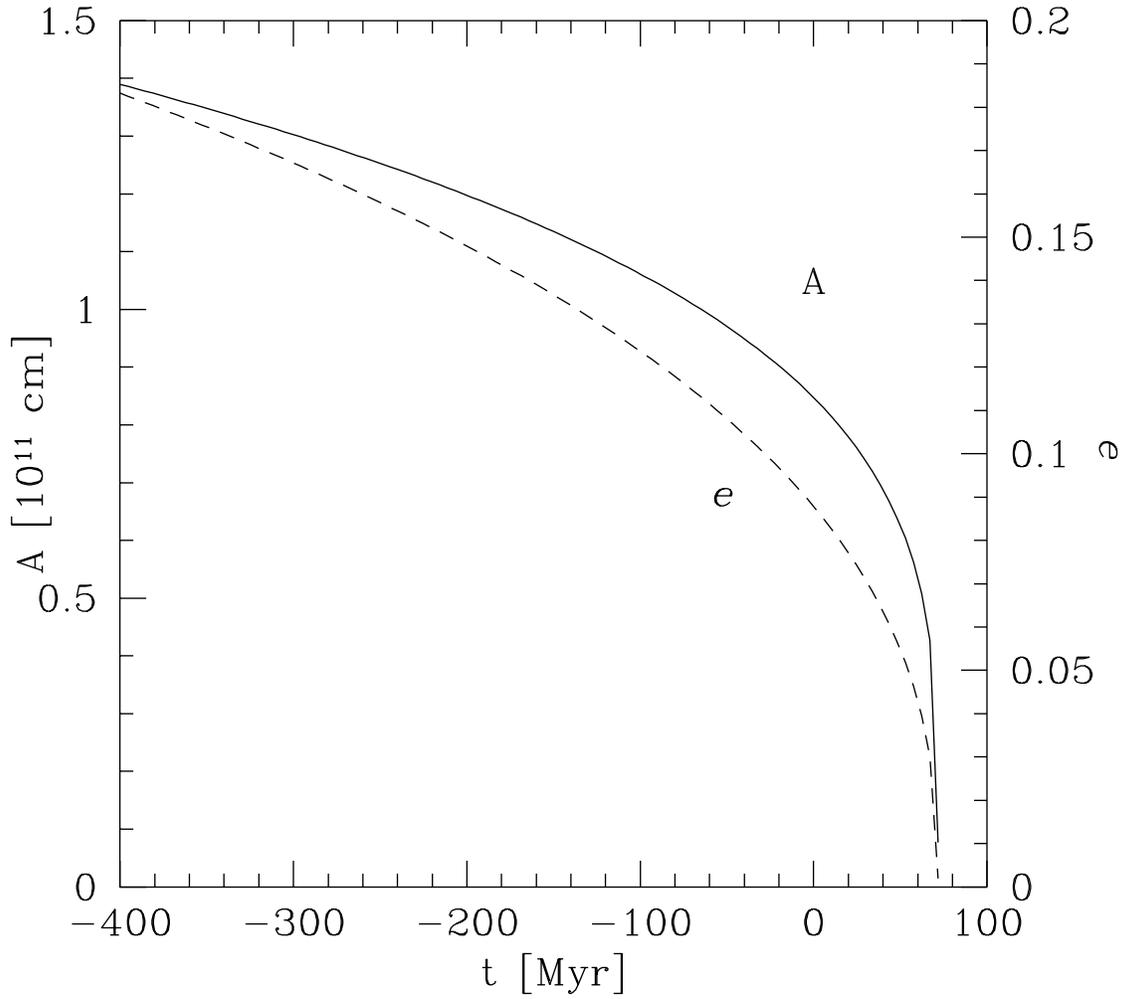,width=6in}}
\caption{ \small \sf
The evolution of the binary ``separation" $A$ defined as the sum of the semi-major axes of both pulsars and the orbital eccentricity $\ee$ due to the emission of gravitational waves as a function of time. This was achieved by integrating the orbital parameter evolution equations \citep{TK} both forward and backward from the current configuration with $A = 8.8\times 10^{10}\mathrm{cm}$ and $\ee =0.088$.
 }
 \label{fig:evol}
 \end{figure}

At present the orbital evolution of the two
pulsar's system is determined by emission of gravitational
radiation. This causes a spiraling in of the binary orbit. The
orbital evolution is well known and can be integrated forwards and
backwards in time according to the formulae given, for example, by
Shapiro and Teukolsy (1983). Fig. \ref{fig:evol} depicts the
evolution of the orbit due to gravitational radiation emission. An
interesting feature of the evolution is that the eccentricity
decays with a faster rate than the orbital separation. Thus an
eccentric orbit becomes, first circular and then it decays.

Fig. \ref{fig:evol} shows that the system will merge within
85~Myrs. This makes it the shortest lived known binary pulsar
system. This short life time has led to a revision upwards of the
estimated rate of binary neutron star mergers in the Galaxy by an
order of magnitude \citep{Burgay2003,Kalogera2004}.
Fig. \ref{fig:evol} also shows the integration backward in time.
This enables us to determine the parameters of the system when it
was born. We find that 50~Myr ago, when the pulsar B was born,
the ellipticity was slightly larger (${\it e}_i \sim 0.11$) and
the semi-major axis was $\sim 10^{11}$cm. Thus we conclude that
the system was born with a low ellipticity and the present low
ellipticity is not due to gravitational radiation decay of the
ellipticity. Instead, it resulted from the orbital parameters of the
system after the second core collapse. Note that the values are not that
different at 200~Myr. At that time the ellipticity was $0.14$ and
the separation was $1.2 \times 10^{11}$cm.

We come now to our main consideration. We ask how can the system
form with both a low eccentricity and a small velocity? We show
that this implies that the mass loss during the second core
collapse was minimal. We assume that prior to the second core
collapse, tidal interaction between the progenitor and the neutron
star has led to a circular orbit. We consider the influence of
mass ejection during this core collapse event on the orbital
motion. We begin with a simple analytic toy model in which we
ignore the eccentricity and demand that the binary neutron star
system moves in circular orbit.

\subsection{A Toy Model}
Consider two masses $m_i$ and $m_2$ in a circular motion with a
separation $A_i$, before the NS formation. The star $i$ explodes
and sheds some of its mass. The remaining mass is $m_f$. Within
the context of \pul $m_f=m_B$ and $m_2=m_A$. We assume that the
explosion is non-spherical and as a result the star attains a kick
velocity $v_k$ in opposite direction to its absolute velocity.
Prior to the explosion, the relative velocity of $m_i$ relative to
$m_2$ is:
\begin{equation}
\label{eq:vc}
 v_{c,i}^2 = {G (m_2+m_i) \over A_i} .
\end{equation}

In order for the new orbit to be circular, the relative velocity
of $m_f$ relative to $m_2$ should be:
\begin{equation}
 v_{c,f}^2 = {G (m_2+m_f) \over A_i} .
\end{equation}
This leads to a condition for $v_k$:
\begin{equation}
\sqrt{G (m_2+m_i) \over A_i} - v_k = \sqrt{G (m_2+m_f) \over A_i}.
\end{equation}
The mass loss and the kick lead to a  center of mass velocity:
\begin{equation}
v_{cm} = {(m_i-m_f)m_2 \over (m_f+m_2)(m_i+m2)} v_{c,i}^2 + {m_f
\over m_f+m_2} v_k
\end{equation}

Defining: $q \equiv m_2/m_f$;   $f \equiv m_f / m_i$ and  $\zeta
\equiv v_{cm} / v_{c,f}$ we get:
\begin{equation}
\zeta =
  \frac{1}{{\sqrt{f}}\,{\sqrt{1 + q}}\,{\sqrt{1 + f\,q}}} -  \frac{1}{1 + q}  ,
\end{equation}
from which it is self evident that with $q \approx 1$, in order
that $\zeta$ should be small (small CM velocity) $f$ must be close
to unity. The results can be understood in the following way. A
spherical mass ejection gives a kick to the CM of the remaining
system direction as the kick given to the CM. The kick is of the
order of the circular velocity times the fraction of the ejected
mass divided by the total remaining mass. Now, in order that the
system remains in a circular orbit the ejected mass should be
ejected non spherically in such a way that it gives a kick to
$m_f$, lowering its velocity relative to $m_2$. This kick should
of the order of the circular velocity of the initial system if the
mass loss is significant. However, this adds a kick to the CM of
the remaining system. This additional kick is in the same
direction as the first one, leading essentially to a CM velocity
which is of the order of the initial circular velocity.

\subsection{The realistic eccentric case}

We consider the same system again, with two masses $m_i$ and $m_2$
moving on a circular orbit before the core collapse.  Now the
system attains an elliptic orbit after the SN. $A$ will now be the sum of
the orbits semi-major axes, or separation in case the pre-collapse circular orbit.
The analysis uses the same
principles as \cite{Hills83,Kalogera96,FryerKalogera97,Wex00}.
Here, we will obtain equations for the unknown initial progenitor
mass $m_i$, the unknown initial separation
$A_i$ and the kick velocity $v_k$, given the current masses,
orbital configuration and binary CM velocity.

The relative velocity of $m_i$ before and after the SN are ${\bf
v}_{rel,i} $ and $ {\bf v}_{rel,f} = {\bf v}_{rel,i} + {\bf v_k}$,
where ${\bf v}_k$ is the kick velocity. The condition of circular
motion before the core collapse gives: $v_{rel,i}^2 =v_{c,i}$, where
$v_{c,i}$ is given by Eq.~\ref{eq:vc}. Defining $\theta$ as the angle
between $v_k$ and $v_{rel,i}$ and the orthogonal angle $\phi$ to
be zero if $v_k$ lies within the plane defined by the momentary
separation between the two masses, ${\bf r}$,  and $v_{rel,i}$,
and pointing outwards from ${\bf r}$, we have,
\begin{equation}
v_{rel,f} = \sqrt{ ({v}_{rel,i}+v_k \cos \theta)^2 + (vk \sin
\theta)^2}.
\end{equation}

The energy of the system after the core collapse is:
\begin{equation}
E = - G {m_f m_2 \over A_i} + {1\over 2} \mu \left[
({v}_{rel,i}+v_k \cos \theta)^2 + (v_k \sin \theta)^2\right].
\end{equation}
where $\mu = m_f m_2 / (m_f + m_2)$ is the reduced mass (after the
core collapse)   and $A_i$ is the separation just prior to the core collapse
(because before the collapse, the orbit was circular). The angular
momentum and the eccentricity $\ee$ after the core collapse are:
\begin{equation}
L_f = A_i \mu \sqrt{ ({v}_{rel,i}+v_k \cos \theta)^2 + (vk \sin
\theta \sin \phi)^2}
\end{equation}
and
\begin{equation}
\label{eq:ee}
\ee^2 = 1+ { 2 L_f^2 E_f \over G^2 (m_f +m_2) \mu^3}
\end{equation}
The new separation $A_f$, satisfies:
\begin{equation}
\label{eq:af}
A_f =  - {G (m_f + m_2) \mu  \over 4 E_f}
\end{equation}
Finally,  the velocity of the center of mass, ${\bf v}_{cm}$ after
the core collapse  is given by:
\begin{equation}
{\bf v}_{cm} = \left( {m_i-m_f \over m_f+m_2}\right) \left({m_2
\over m_i +m_2}\right) {\bf v}_{rel,i} + {m_f \over m_f+m2} {\bf
v}_k .
\end{equation}
Using our definitions of angles, we have:
\begin{equation}
v_{cm}^2 = {m_i-m_f \over m_f + m_2} \left\{ \left[ \left({m_2
\over m_i +m_2}\right) v_{rel,i} - \left( m_f \over \ {m_i-m_f}
\right) v_k \cos \theta \right]^2  + \left[\left( m_f \over \
{m_i-m_f} \right) v_k \sin \theta  \right]^2   \right\}
\end{equation}

We define an equivalent circular velocity for the binary after the
core collapse:
\begin{equation}
 v_{c,f}^2 = { G (m_f + m_2) \over A_f}.
\end{equation}
This is a fiducial quantity that we will use as a measure of the
various velocities in the system.  Defining $k \equiv v_k /
v_{c,f}$, and $\alpha \equiv A_f/A_i$, and using the previous
definitions of $f$, $q$, and $\zeta$ we obtain three equations for
the eccentricity, the semi-major axis and the CM velocity:
\begin{eqnarray}
\ee^2 & = & {1\over {\alpha^2  f^2  {\left( 1 + q \right) }^2}}
\left( f  k^2  \left( 1 + q \right)  - \alpha  \left( f \left( 2 +
q \right) -1  \right) + {2\sqrt{\alpha}}  f k {\sqrt{1 + q}}
{\sqrt{\frac{1}{f} + q}}  \cos \theta
\right) \nonumber \\
& & \times \left( \alpha \left( 1 +  f  q \right)  + {2\sqrt{\alpha}} f k
{\sqrt{1 + q}} {\sqrt{\frac{1}{f} + q}}
   \cos \theta +  f  k^2  \left( 1 + q \right)   {\cos \theta}^2
   + \right.\\
& & \left.  f  k^2  \left( 1 + q \right)   {\sin \theta}^2 {\sin
(\phi )}^2
 \right) , \nonumber
 \end{eqnarray}
\begin{equation}
1= \frac{   f  \left( 1 + q \right) } { \alpha  \left( f  \left( 2
+ q \right)-1 \right)  -\left( f k^2 \left( 1 + q \right)  \right)
-   {2\sqrt{\alpha}}  f k {\sqrt{1 + q}} {\sqrt{\frac{1}{f} + q}}
\cos \theta}
\end{equation}
and
\begin{equation}
{\zeta }^2 = \frac{\alpha  {\left( f-1 \right) }^2  q^2 + f k^2
\left( 1 + q \right)   \left( 1 + f  q \right)  +  {2\sqrt{\alpha}}
\left(  f-1 \right)
  f  k  q  {\sqrt{1 + q}}  {\sqrt{\frac{1}{f} + q}}   \cos \theta}{ f
  {\left( 1 + q \right) }^3  \left( 1 + f  q \right) }
\end{equation}

For a given $\zeta$, $\ee$, $\theta$ and $\phi$ a solution to the
three equations can be found numerically for $f$, $k$, and $\alpha$.
Then, the $\theta$ and $\phi$ phase space is searched to find the
minimum value of $f$ possible. This gives the maximal mass that
can be ejected for a given CM velocity.

\begin{figure}
\hskip -0.5cm
\center{\epsfig{file=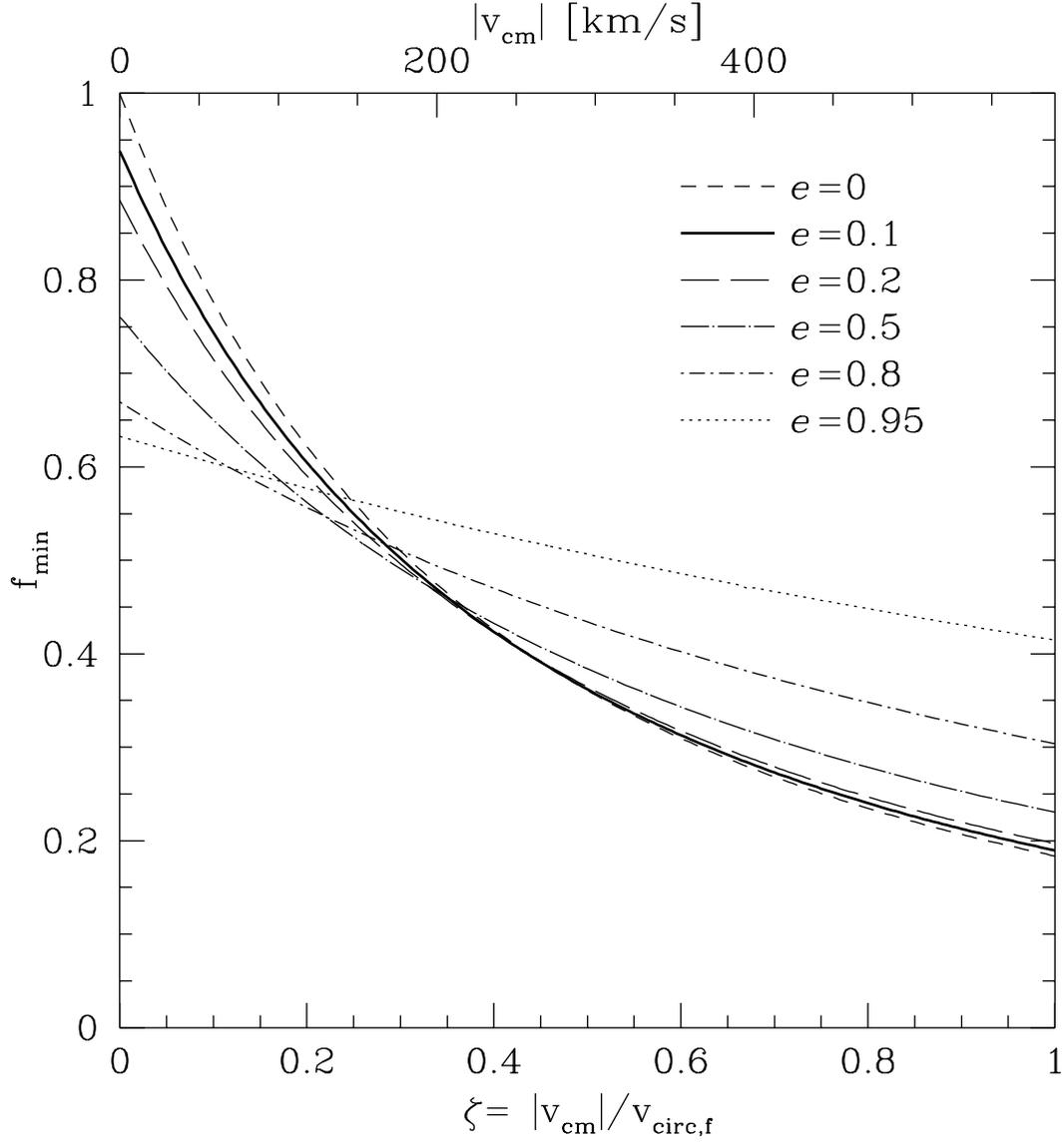,width=6in}}
\caption{ \small \sf
The minimal fraction of mass $f \equiv m_f / m_i$ left after the core collapse in units of the initial mass $m_i$ of the primary, plotted as a function of the kick velocity $|{\bf v_{cm}}|$, or in a dimensionless velocity $\zeta \equiv |{\bf v_{cm}}|/v_{circ,f}$. The different graphs are calculated for different eccentricities, with $\ee \approx 0.1$ being the preferred value at $t \sim - 50$~Myr, the rough age based on the spin down age. $\zeta$ is expected to be $\sim 0.02$. Otherwise, the binary would have been far from the galactic plane, unless by chance it was kicked very close to the plane.
 }
 \label{fig:fzeta}
 \end{figure}

Fig. \ref{fig:fzeta} depicts, for various initial eccentricities,
the minimal $f$ value as a function of $\zeta$. The integration
backwards of the orbit to the time of the core collapse, about 50~Myr ago
shows that  just after the core collapse  the orbital separation was  $\sim
10^{11}$cm  and the eccentricity, was around $0.11$. The
corresponding circular velocity was 590~km/sec. Given that the
current CM motion is most probably of order 15~km/sec, $\zeta <
0.025$ is a reasonable upper limit. Given the eccentricity 0.1, this value of $\zeta$
implies that $f > 0.88$ or that the mass loss during the
formation of the second neutron star formation was probably of order $\sim 0.15 M_\sun$.
For $\zeta < 0.25$ (our 95\% confidence limit), we find that $f_{min} = 0.55$, thus,
the star shed less than $1.0~M_{\sun}$ at 95\% confidence during the core collapse event.
The progenitor was most probably less massive than $2.3 M_{\sun}$,
with $m_i \sim 1.4 M_\sun$ being more likely.

\section{Conclusions and Implications}
\label{sec:concl}

The binary B1913+16 was studied for almost 3 decades since its
discovery \citep{Hulse75}. The ample data could then be used by
\cite{Wex00} to place very interesting limits on this system. For
example, it was found that the natal kick must have been directed
almost perpendicular to the spin axis of the neutron star
progenitor. On the other hand, the rather large post-SN
eccentricity implies that the pre-collapse orbital separation
($1.8-4.6~R_\sun$) or the mass of the progenitor ($4-32~M_\sun$) cannot be
tightly constrained.

The system \pul is different from previously detected binary pulsar systems in that it has
a very low eccentricity, and it was small also after the collapse.
Thus, the orbital separation, for example, can be constrained to have been $A_i = 1.0 \pm 0.1 \times 10^{11}
cm = 1.45 \pm 0.15~R_{\sun}$ during the collapse. As the system \pul was only recently
discovered,  the proper motion could not be detected yet.
Therefore, in order to get meaningful constraints on the kinematics of the SN explosion, \cite{DvdH04} and \cite{WK04} constrained in their analyses the progenitor mass to be consistent with the evolution of He stars in tight binaries.

Statistically,  however, it is possible to place limits on the
center of mass motion of the system, at different confidence
limits, using the fact that it is within about 50~pc from the
galactic plane. In particular, $v_{cm} \sim 15$~km/s and it is
less than $150$~km/s at about 95\% confidence. This implies that
the ejected mass during the collapse is likely to be less than
$\sim 0.15~M_{\sun}$, while at 95\% confidence, it is less than
$\sim 1.0~M_\sun$.

Since current evolutionary models for Helium stars require a minimum mass of 2.1 to 2.8$M_{\sun}$ to form a NS \citep[and references therein]{TvdH}, there seems to be two reasonable possibilities.  

In the first possibility, \pulB originated in a SN by a progenitor Helium star with $M_\sun \gtrsim 2.1 M_\sun$. This case requires the systemic kick to have been large but within the galactic plane. This could have occurred with a low probability of $\lesssim 5\%$. Since He stars which lose mass only through winds generally have somewhat higher masses ranging between $2.3$ to $3.5~M_\sun$
\citep{Woosley95}, a possible interpretation is  that the  progenitor star lost most of its envelope
through interaction with its companion. This is also reasonable considering the small separation.  Since most of the lost mass could not have
accreted onto the companion, it was probably lost  through a
common envelope phase, at which point the companion \pulA was spun
up and its magnetic field was suppressed by accretion. This
accretion phase ended around 200~Myr ago.

Because so little of the star was left after the mass loss, it is
possible that even the Helium mantle was removed, such that the
resulting supernova, if it were a supernova, was of Type Ic.

A second interpretation of the above results follows the more probable mass loss.
We expect  the mass loss in the cataclysmic event to have been
$\lesssim 0.15 M_\sun$, given the small systemic velocity. Moreover, we note that just the $\nu$ burst from a core
collapse should reduce the mass by the same order, with a typical value being $\sim 10^{53} erg/c^2 \sim 0.05
M_\sun$. In other words,  the collapse may have been  of a naked core.
Such a collapse does not eject any appreciable mass, but it sheds
enough ``mass-energy" in the form of $\nu$'s to explain the small
eccentricity. In such a case, no kick velocity is expected, and
we can determine the $\nu$ burst energy from the inferred eccentricity
after the collapse. Namely,  $E_\nu  \approx e (1+q)/[(1+e (1+q)]
m_B c^2 = 0.228^{+0.025}_{-0.015} M_\odot c^2 \approx 4.2 \times 10^{53} erg$ (assuming the collapse took place $50^{+50}_{-25}$~Myr ago). If some mass is ejected, $E_\nu$ will of course be smaller but the total mass-energy lost will be the same. Thus, under the kinematically more probably option, this collapse did not proceed along the common core
collapse scenario. Instead the WD progenitor, with a mass of $M_B+E_\nu/c^2 = 1.478^{+0.025}_{-0.015} M_\sun$ was supported by thermal pressure as well as degeneracy and that the collapse was induced
due to the loss of this thermal pressure as the progenitor was
cooling. It could have also ensued other internal transitions such as the process of slow $e$-capture which reduces the Chandrasekhar limit \citep{Finzi}.

At present we cannot, however, rule out the kinematically less probable possibility
that the system does have a large CM velocity and that this
velocity is within the Galactic plane. This possibility implies a He star progenitor for \pulB. The kinematically more favorable possibility implies that the pulsar is the remnant of the naked collapse of a WD.
 
The future detection within a few years of a
proper motion could constrain the center of mass velocity much
more, after an  additional velocity component would be determined.  A constrained peculiar motion will probably
solidify this extreme case of a low mass collapse and resolve the puzzle on the origin of the object.
A useful measurement in this respect, is of the inclination of the
pulsar orbit relative to the galactic plane. This could be achieved
through measurements of the polarization behavior and of the
relativistic spin precession.

The authors wish to thank Ramesh Narayan and members of the
Israeli center for High Energy Astrophysics, D. Eichler, M.
Milgrom, E. Waxman and V. Usov  for helpful comments. The research
was supported by an ISF grant.

 \end{document}